\documentclass{elsart}[12pt]%
\usepackage{epsfig,subfigure,amsmath,url,amssymb}
\newcommand{\aff}[2]{Dipartimento di Fisica dell'Universit\`a #1 e Sezione INFN, #2, Italy.}
\newcommand{\affd}[1]{Dipartimento di Fisica dell'Universit\`a e Sezione INFN, #1, Italy.}
\newcommand{\be}{\begin{equation}}
\newcommand{\ee}{\end{equation}}
\newcommand{\bea}{\begin{eqnarray}}
\newcommand{\eea}{\end{eqnarray}}
\newcommand{\bc}{\begin{center}}
\newcommand{\ec}{\end{center}}
\newcommand{\bt}{\begin{tabular}}
\newcommand{\et}{\end{tabular}}
\newcommand{\bfig}{\begin{figure}}
\newcommand{\efig}{\end{figure}}
\newcommand{\bi}{\begin{itemize}}
\newcommand{\ei}{\end{itemize}}
\newcommand{\bleft}{\begin{flushleft}}
\newcommand{\eleft}{\end{flushleft}}
\newcommand{\bright}{\begin{flushright}}
\newcommand{\eright}{\end{flushright}}
\newcommand{\bpage}{\begin{minipage}}
\newcommand{\epage}{\end{minipage}}

\newcommand{\pip}{\ensuremath{\pi^+\,}}
\newcommand{\pim}{\ensuremath{\pi^-\,}}
\newcommand{\piz}{\ensuremath{\pi^0\,}}
\newcommand{\Ks}{\ensuremath{K_S \,}}
\newcommand{\Kl}{\ensuremath{K_L\,}}

\newcommand{\Eta}{\ensuremath{\eta\,}}
\newcommand{\etap}{\ensuremath{\eta'\,}}

\renewcommand{\to}{\ensuremath{\rightarrow}}
\newcommand{\cm}{\ensuremath{\,{\rm cm}}}

\newcommand{\s}{\ensuremath{\,{\rm s}}}

\newcommand{\pb}{\ensuremath{\,{\rm pb}}}
\newcommand{\pbinv}{\ensuremath{\,{\rm pb}^{-1}}}

\newcommand{\MeV}{\ensuremath{\,{\rm MeV}}}
\newcommand{\fikskl}{\ensuremath{\phi\rightarrow\Ks\Kl}}
\newcommand{\fipippimpiz}{\ensuremath{\phi\rightarrow\pi^+\pi^-\pi^0}}
\newcommand{\fietag}{\ensuremath{\phi\rightarrow\eta\gamma\;}}

\newcommand{\etapippimpiz}{\ensuremath{\eta\rightarrow\pip\pim\piz}}

\newcommand{\etatrepi}{\ensuremath{\eta\rightarrow3\pi}}
\begin{document}
\begin{frontmatter}
\title{
Determination of $\eta\to\pi^+\pi^-\pi^0 $ Dalitz Plot slopes and asymmetries with the KLOE detector. }

\collab{The KLOE Collaboration} 
\author[Na]{F.~Ambrosino\thanksref{*}},
\author[Frascati]{A.~Antonelli},
\author[Frascati]{M.~Antonelli},
\author[Frascati]{F.~Archilli},
\author[Roma3]{C.~Bacci},
\author[Karlsruhe]{P.~Beltrame},
\author[Frascati]{G.~Bencivenni},
\author[Frascati]{S.~Bertolucci},
\author[Roma1]{C.~Bini},
\author[Frascati]{C.~Bloise},
\author[Roma3]{S.~Bocchetta},
\author[Roma1]{V.~Bocci},
\author[Frascati]{F.~Bossi},
\author[Roma3]{P.~Branchini},
\author[Roma1]{R.~Caloi},
\author[Frascati]{P.~Campana},
\author[Frascati]{G.~Capon},
\author[Frascati]{T.~Capussela\thanksref{*}},
\author[Roma3]{F.~Ceradini},
\author[Frascati]{S.~Chi},
\author[Na]{G.~Chiefari},
\author[Frascati]{P.~Ciambrone},
\author[Frascati]{E.~De~Lucia},
\author[Roma1]{A.~De~Santis},
\author[Frascati]{P.~De~Simone},
\author[Roma1]{G.~De~Zorzi},
\author[Karlsruhe]{A.~Denig},
\author[Roma1]{A.~Di~Domenico},
\author[Na]{C.~Di~Donato},
\author[Pisa]{S.~Di~Falco},
\author[Roma3]{B.~Di~Micco},
\author[Na]{A.~Doria},
\author[Frascati]{M.~Dreucci},
\author[Frascati]{G.~Felici},
\author[Frascati]{A.~Ferrari},
\author[Frascati]{M.~L.~Ferrer},
\author[Frascati]{G.~Finocchiaro},
\author[Roma1]{S.~Fiore},
\author[Frascati]{C.~Forti},
\author[Roma1]{P.~Franzini},
\author[Frascati]{C.~Gatti},
\author[Roma1]{P.~Gauzzi},
\author[Frascati]{S.~Giovannella},
\author[Lecce]{E.~Gorini},
\author[Roma3]{E.~Graziani},
\author[Pisa]{M.~Incagli},
\author[Karlsruhe]{W.~Kluge},
\author[Moscow]{V.~Kulikov},
\author[Roma1]{F.~Lacava},
\author[Frascati]{G.~Lanfranchi},
\author[Frascati,StonyBrook]{J.~Lee-Franzini},
\author[Karlsruhe]{D.~Leone},
\author[Frascati]{M.~Martini},
\author[Na]{P.~Massarotti},
\author[Frascati]{W.~Mei},
\author[Na]{S.~Meola},
\author[Frascati]{S.~Miscetti},
\author[Frascati]{M.~Moulson},
\author[Frascati]{S.~M\"uller},
\author[Frascati]{F.~Murtas},
\author[Na]{M.~Napolitano},
\author[Roma3]{F.~Nguyen},
\author[Frascati]{M.~Palutan},
\author[Roma1]{E.~Pasqualucci},
\author[Roma3]{A.~Passeri},
\author[Frascati,Energ]{V.~Patera},
\author[Na]{F.~Perfetto\thanksref{*}},
\author[Lecce]{M.~Primavera},
\author[Frascati]{P.~Santangelo},
\author[Na]{G.~Saracino},
\author[Frascati]{B.~Sciascia},
\author[Frascati,Energ]{A.~Sciubba},
\author[Pisa]{F.~Scuri},
\author[Frascati]{I.~Sfiligoi},
\author[Frascati]{T.~Spadaro},
\author[Roma1]{M.~Testa},
\author[Roma3]{L.~Tortora},
\author[Roma1]{P.~Valente},
\author[Karlsruhe]{B.~Valeriani},
\author[Frascati]{G.~Venanzoni},
\author[Frascati]{R.Versaci},
\author[Frascati,Beijing]{G.~Xu}
\address[Frascati]{Laboratori Nazionali di Frascati dell'INFN, 
Frascati, Italy.}
\address[Karlsruhe]{Institut f\"ur Experimentelle Kernphysik, 
Universit\"at Karlsruhe, Germany.}
\address[Lecce]{\affd{Lecce}}
\address[Na]{Dipartimento di Scienze Fisiche dell'Universit\`a 
``Federico II'' e Sezione INFN,
Napoli, Italy}
\address[Pisa]{\affd{Pisa}}
\address[Energ]{Dipartimento di Energetica dell'Universit\`a 
``La Sapienza'', Roma, Italy.}
\address[Roma1]{\aff{``La Sapienza''}{Roma}}
\address[Roma2]{\aff{``Tor Vergata''}{Roma}}
\address[Roma3]{\aff{``Roma Tre''}{Roma}}
\address[StonyBrook]{Physics Department, State University of New 
York at Stony Brook, USA.}
\address[Beijing]{Permanent address: Institute of High Energy 
Physics of Academica Sinica,  Beijing, China.}
\address[Moscow]{Permanent address: Institute for Theoretical 
and Experimental Physics, Moscow, Russia.}
\thanks[*]{Corresponding authors.\\
           {\it E-mail address:} Fabio.Ambrosino@na.infn.it (F. Ambrosino)\\
           Tiziana.Capussela@lnf.infn.it (T. Capussela)\\
           Francesco.Perfetto@na.infn.it (F. Perfetto).}

\begin{abstract}
We have studied, with the KLOE detector at the DA$\Phi$NE $\Phi$-Factory, the dynamics of
the decay \etapippimpiz using data from
the radiative $\Phi \to \eta \gamma $ decay
 for an integrated luminosity $L =
450 \pbinv$. 
From a fit to the Dalitz plot density distribution we obtain a precise
measurement of the slope parameters. This should allow to improve the knowledge of
the decay amplitude which is sensitive to the u-d quark
mass difference. 
We also present new best results on the C-violating asymmetries in the
\etapippimpiz decay.
\end{abstract}
\end{frontmatter}

\def\ifm#1{\relax\ifmmode#1\else$#1$\fi}
\def\eps{\ifm{\epsilon}} \def\epm{\ifm{e^+e^-}}
\def\rep{\ifm{\Re(\eps'/\eps)}}  \def\imp{\ifm{\Im(\eps'/\eps)}}  
\def\DAF{DA$\Phi$NE}  \def\sig{\ifm{\sigma}}
\def\gam{\ifm{\gamma}} \def\to{\ifm{\rightarrow}}
\def\pip{\ifm{\pi^+}} \def\pim{\ifm{\pi^-}}
\def\po{\ifm{\pi^0}} 
\def\pic{\ifm{\pi^+\pi^-}} \def\pio{\ifm{\pi^0\pi^0}} 
\def\ks{\ifm{K_S}} \def\kl{\ifm{K_L}} \def\kls{\ifm{K_{L,\,S}}} 
\def\ksl{\ifm{K_S,\ K_L}} \def\ko{\ifm{K^0}}
\def\K{\ifm{K}} \def\LK{\ifm{L_K}}
\def\Kb{\ifm{\rlap{\kern.3em\raise1.9ex\hbox to.6em{\hrulefill}} K}}
\def\ab{\ifm{\sim}}  \def\x{\ifm{\times}}
\def\ff{$\phi$--factory}
\def\sta#1{\ifm{|\,#1\,\rangle}} 
\def\amp#1,#2,{\ifm{\langle#1|#2\rangle}}
\def\kob{\ifm{\Kb\vphantom{K}^0}}
\def\f{\ifm{\phi}}   \def\pb{{\bf p}}
\def\L{\ifm{{\cal L}}}  \def\R{\ifm{{\cal R}}}
\def\up#1{$^{#1}$}  \def\dn#1{$_{#1}$}
\def\etal{{\it et al.}}
\def\BR{{\rm BR}}
\def\radl{\ifm{X_0}}
\def\deg{\ifm{^\circ}} 
\def\th{\ifm{\theta}}
\def\To{\ifm{\Rightarrow}}
\def\ot{\ifm{\leftarrow}}
\def\fo{\ifm{f_0}} \def\epe{\ifm{\eps'/\eps}}
\def\pbrn{ {\rm pb}}  \def\cm{ {\rm cm}}
\def\mub{\ifm{\mu{\rm b}}} \def\s{ {\rm s}}
\def\RR{\ifm{{\cal R}^\pm/{\cal R}^0}}
\def\dt{ \ifm{{\rm d}t} } \def\dy{ {\rm d}y } \def\pbrn{ {\rm pb}}
\def\kp{\ifm{K^+}} \def\km{\ifm{K^-}}
\def\kkb{\ifm{\ko\kob}} 
\def\epe{\ifm{\eps'/\eps}}
\def\ppc{\ifm{\pi^+\pi^-}}
\def\ppo{\ifm{\pi^0\pi^0}}
\def\pppco{\ifm{\pi^+\pi^-\pi^0}}
\def\pppo{\ifm{\pi^0\pi^0\pi^0}}
\def\vare{\ifm{\varepsilon}}
\def\etap{\ifm{\eta'}}

\section{Introduction}
The decay of the isoscalar $\eta$ into three pions occurs through 
isospin violation and thus is sensitive to the up-down
quark mass difference.
The electromagnetic corrections to the decay are small (Sutherland's
theorem \cite{Sutherland}) and do not affect significantly the rate or the Dalitz plot density \cite{BaKaWy96}.\\
Neglecting electromagnetic corrections, the decay amplitude is given
by \cite{BijGa02}:

\begin{equation}
A(s,t,u)=\frac{1}{Q^{2}}\frac{m_{K}^{2}}{m_{\pi}^{2}}
\left(m_{\pi}^{2}-m_{K}^{2}\right)\frac{M(s,t,u)}{3 \sqrt{3} F^{2}_{\pi}}
\end{equation}
\noindent
where 
\begin{equation}
Q^{2}\equiv\frac{m_{s}^{2}-\hat{m}^{2}}{m_{d}^{2}-m_{u}^{2}}
\end{equation}
is a combination of quark masses and
$\hat{m}=\frac{1}{2}\left(m_{u}+m_{d}\right)$ is the average u, d
quark mass, $F_{\pi} = 92.4$ MeV is the pion decay constant and
$M(s,t,u)$ contains all the dependence of the amplitude on the
Mandelstam invariants. Since the decay rate is proportional to
$Q^{-4}$,
\begin{equation}
 \Gamma\left(\etapippimpiz\right) \propto \vert A\vert^{2} \propto Q^{-4} 
\end{equation}
the transition \etatrepi~  is, in principle, a very
sensitive probe to determine $Q$. A theoretical estimate of Q is obtained from
the Dashen theorem\footnote{The Dashen theorem states that in the
chiral limit the charged kaon and pion
electromagnetic mass shifts are the same:
$\left(m_{\pip}^{2}-m_{\piz}^{2}\right)_{em}
=\left(m_{K_{+}}^{2}-m_{K^{0}}^{2}\right)_{em}$} \cite{Dashen}  
according to which
\begin{equation}
Q^{2}_{Dashen}=\frac{m_{K}^{2}}{m_{\pi}^{2}}\frac{m_{K}^{2}-m_{\pi}^{2}}
{m_{K^{0}}^{2}-m_{K^{+}}^{2} + m_{\pip}^{2}-m_{\piz}^{2}}= ( 24.1)^{2}. 
\end{equation}
\noindent 
Of course, in order to extract the quark mass ratio from the decay
width, one needs an accurate description of
$M(s,t,u)$. \\
At lowest order 
\begin{equation}
\label{EqLO}
M(s,t,u) =\frac{3 s-4 m_{\pi}^{2}}{m_{\eta}^{2}-m_{\pi}^{2}}.
\end{equation}
\noindent 
a well known result, based on Current Algebra. 
Integrating eq.(\ref{EqLO}) over the phase space
and using the quark masses as estimated by Leutwyler \cite{Leut} one obtains 
the following prediction for the decay rate \cite{BijGa02} :
\begin{equation}
\Gamma^{theo}\left(\etapippimpiz\right)=66\;\textrm{eV}
\label{eq:theo1}
\end{equation}
\noindent
which strongly disagrees with the experimental value \cite{PDG}:
\begin{equation}
\Gamma^{exp}\left(\etapippimpiz\right)= 295 \pm 16 \;\textrm{eV}. 
\end{equation}
\noindent
A one-loop calculation within conventional chiral perturbation theory
($\chi$PT) \cite{GasLut85}, improves  considerably the prediction: 
\begin{equation}
\Gamma^{theo}\left(\etapippimpiz\right)\simeq 167 \pm 50
\;\textrm{eV}. 
\label{eq:theo2}
\end{equation}
\noindent
but is still far from the experimental value.\\
Higher order corrections discussed in the literature \cite{BijGa02}
contribute to
increase the theoretical prediction, but cannot fully account for the
 discrepancy with the data. More recently a good agreement has been
found combining the effective chiral Lagrangian with a non perturbative
scheme based on coupled channels and Bethe Salpeter equation \cite{Borasoy}.

A significant violation of the Dashen theorem 
 could in principle account for the discrepancy; however it should 
 be claimed only after demonstrating  agreement between theory and experiment on
the $M(s,t,u)$ amplitude behaviour over phase space.

The above discussion therefore motivates  a precise measurement of the \etatrepi~
dynamics through the study of the Dalitz plot.\\
The Dalitz plot of the \etapippimpiz decay is described by two kinematic variables.
By convention these are defined in terms of the kinetic energies of the pions
$T_{+}$, $T_{-}$ and $T_{0}$ in the \Eta rest frame:
\begin{eqnarray}
X = \sqrt{3}\frac{T_{+}-T_{-}}{Q_{\eta}} = \frac{\sqrt{3}}{2
M_{\eta}Q_{\eta}} \left(u-t\right),\nonumber\\
Y = \frac{3 T_{0}}{Q_{\eta}}-1 = \frac{3}{2 M_{\eta}Q_{\eta}}
\left[\left((m_{\eta}-m_{\piz}\right)^{2}-s\right]-1,\nonumber\\
Q_{\eta} = m_{\eta}-2 m_{\pip}-m_{\piz},\nonumber\\
\nonumber
\end{eqnarray}
\noindent
where X and Y  respectively vary in the range
$\left[-1,1\right]$ and $\left[-1,0.895\right]$.\\ 
The decay amplitude is then expanded around the center of the Dalitz plot
(X=Y=0) in powers of X and Y as:
\begin{equation}
\vert A(X,Y)\vert^{2} \simeq 1 + aY + bY^{2} + c X + dX^{2} + eXY +....
\label{eq:amp_tra}
\end{equation}
\noindent
and the parameters ($a,b,c,d,e,...$ ) of the expansion are fitted to
the experimental data.
Any odd power of X contributing to $\vert A(X,Y)\vert^{2}$ would imply
 violation of Charge Conjugation.\\ 


\section{The KLOE detector}
Data were collected with the KLOE detector at DA$\Phi$NE \cite{DAFNE},
the Frascati $e^{+} e^{-}$ collider, which operates at a
center of mass energy $\sqrt {s} =M_{\Phi}\sim$1020
MeV/c$^{2}$. The electron and  positron beams collide with a
crossing angle of $\pi - 25$ mrad, resulting in a small momentum
( $p_{\phi}\sim$ 13 MeV/c in the horizontal plane) of the produced $\phi$
mesons.\\
The KLOE detector consists
of a large cylindrical drift chamber (DC), surrounded by a fine
sampling lead-scintillating fibers electromagnetic calorimeter (EMC) 
inserted in a 0.52 T magnetic field..\\
The DC \cite{DC}, 4 m diameter and 3.3 m long, has full stereo
geometry and operates with a gas mixture of 90\% Helium and 10\%
Isobutane.
Momentum resolution is $\frac{\sigma_{p_{T}}}{p_{T}}\leq
0.4\%$. Position resolution in $r - \phi$ is 150 $\mu$m and $\sigma_{z}\sim$ 2
mm. Charged tracks vertices are reconstructed with an accuracy of $\sim$3 mm.\\
The EMC \cite{EMC} is divided into a barrel and two endcaps, for a total
of 88 modules, and covers 98\% of the solid angle.
Arrival times of particles and space positions  of 
the energy deposits are obtained from the signals collected at the two
ends of the calorimeter modules, with a granularity of $\sim$(4.4 x
4.4) cm$^{2}$, for a total of 2240 cells arranged in five
layers. Cells close in time and space are grouped into a calorimeter
cluster. The cluster energy E is the sum of the cell energies, while
the cluster time T and its position R are energy weighted
averages. The respective resolutions are $
\sigma_E/E = 5.7\% / \sqrt{E(GeV)}$
and $\sigma_T = 57 \;\textrm{ps} / \sqrt{E(GeV)} \oplus 100\;
\textrm{ps}$.\\
The KLOE trigger \cite{TRIGGER} is based on the coincidence of at
least two  energy deposits in the EMC, above a threshold that
ranges between 50 and 150 MeV. In order to reduce the trigger rate due
to cosmic rays crossing the detector, events with a large energy
release in the outermost calorimeter planes are vetoed.

\section{Signal selection and efficiency}
At KLOE  $\eta$ mesons are produced in the process $\fietag$, so to
study the dynamics of $\eta\rightarrow\pip\pim\piz$ the final state 
$\pip\pim\gamma\gamma\gamma$ is used to which corresponds a BR value of:
\begin{equation}
BR_{TOT} =
BR\left(\fietag\right)\times
BR\left(\etapippimpiz\right)\times
BR\left(\piz\rightarrow\gamma\gamma\right) 
= 2.9 \cdot 10^{-3}.
\nonumber 
\end{equation}
There is no combinatorial problem in the photon pairing because this
decay chain is characterised by an energetic monochromatic recoil photon, with
$E_{\gamma_{rec}}\sim 363$ \textrm{MeV}, quite well separated from the
softer  photons from $\piz$ decay.

In the data analysis a photon is defined as  a cluster in the EmC with no associated
  track detected in the Drift Chamber (DC) and with $|(T-\frac Rc)|<
  5\sigma_T$; where $T$ is the arrival time on the EmC, $R$ is the
  distance of the cluster from the vertex, $c$ is the light speed 
  and $\sigma_T$ is the time resolution.

\label{Efficiencies}
The events selection is performed through the following steps:
\begin{enumerate}
\item Events are selected starting from a very loose offline
  preselection consisting of a machine
  background filter (FILFO) and an event selection procedure (EVCL)
  that assigns events into categories \cite{EVCL}.
\item One charged vertex is required inside the cylindrical region
  $r<4\;\cm$, $|z|<8 \;\cm$ and 3 photons with $ 21^{\circ}<\theta_{\gamma}<159^{\circ}$ and
  $E_{\gamma}>10$ MeV. The probability of a photon to fragment in more
  than a cluster (splitting) is reduced by requiring the
  opening angle between any pair of photons to be $>18^{\circ}$. 
\item $\sum E_{\gamma} <800\MeV.$ 
\item A constrained kinematic fit is performed imposing 4-momentum
  conservation and $t=\frac {R}{c}$ for each photon, without imposing
  mass constraint both on $\eta$ and $\piz$.\\
  A cut on the $\chi^{2}$ probability is done, P$(\chi^{2})>1$\%.\\
  This fit improves
  significantly the resolution on photon energies.
  The  $\chi^2$ distribution
  of the fit shows a satisfactory Data-MC agreement as shown in fig.\ref{chisq_eta_2}: 
  
\begin{figure}[h]
\begin{center}
\begin{tabular}{cc}
\epsfig{figure=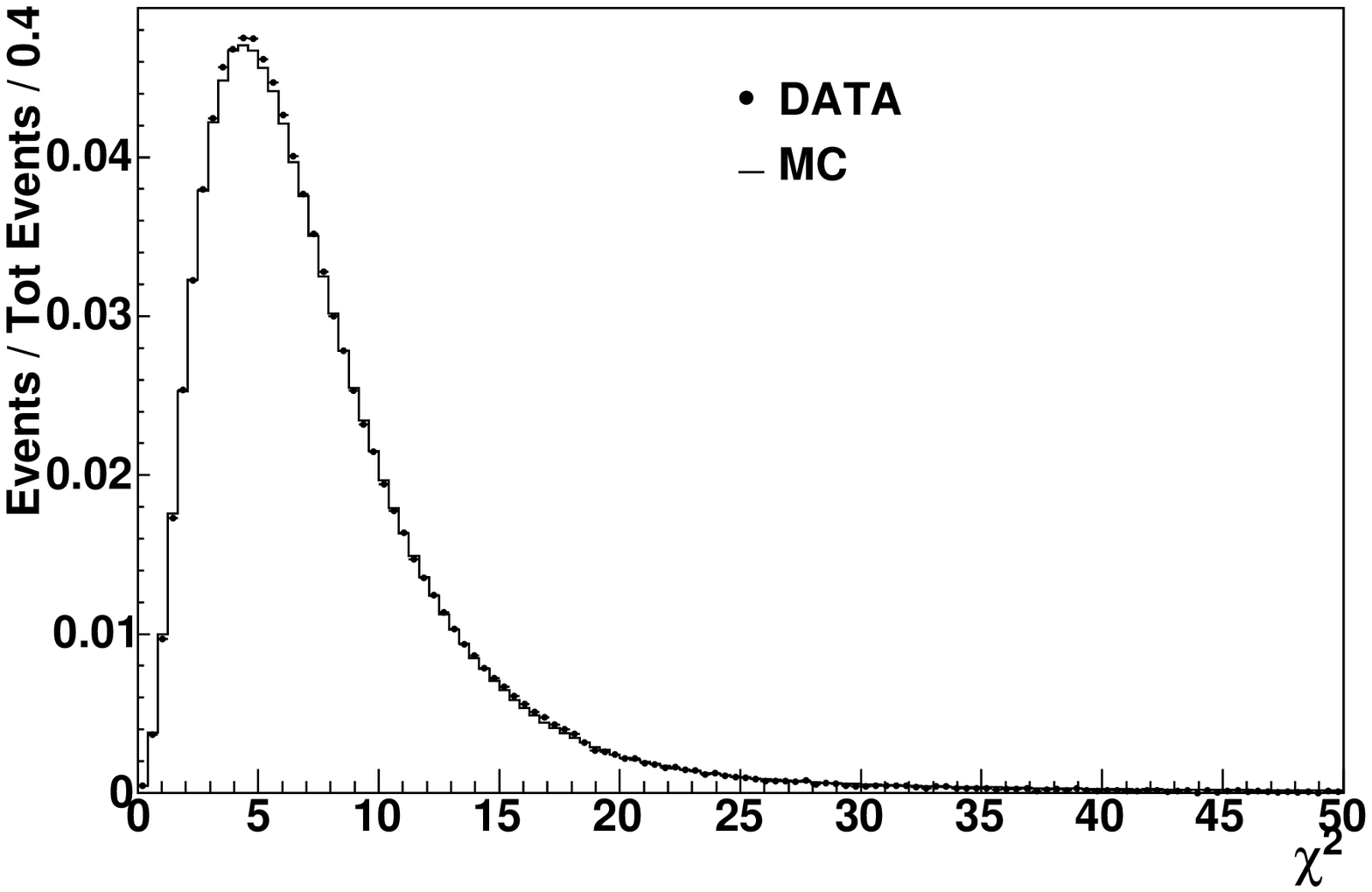, width=80mm}&
\epsfig{figure=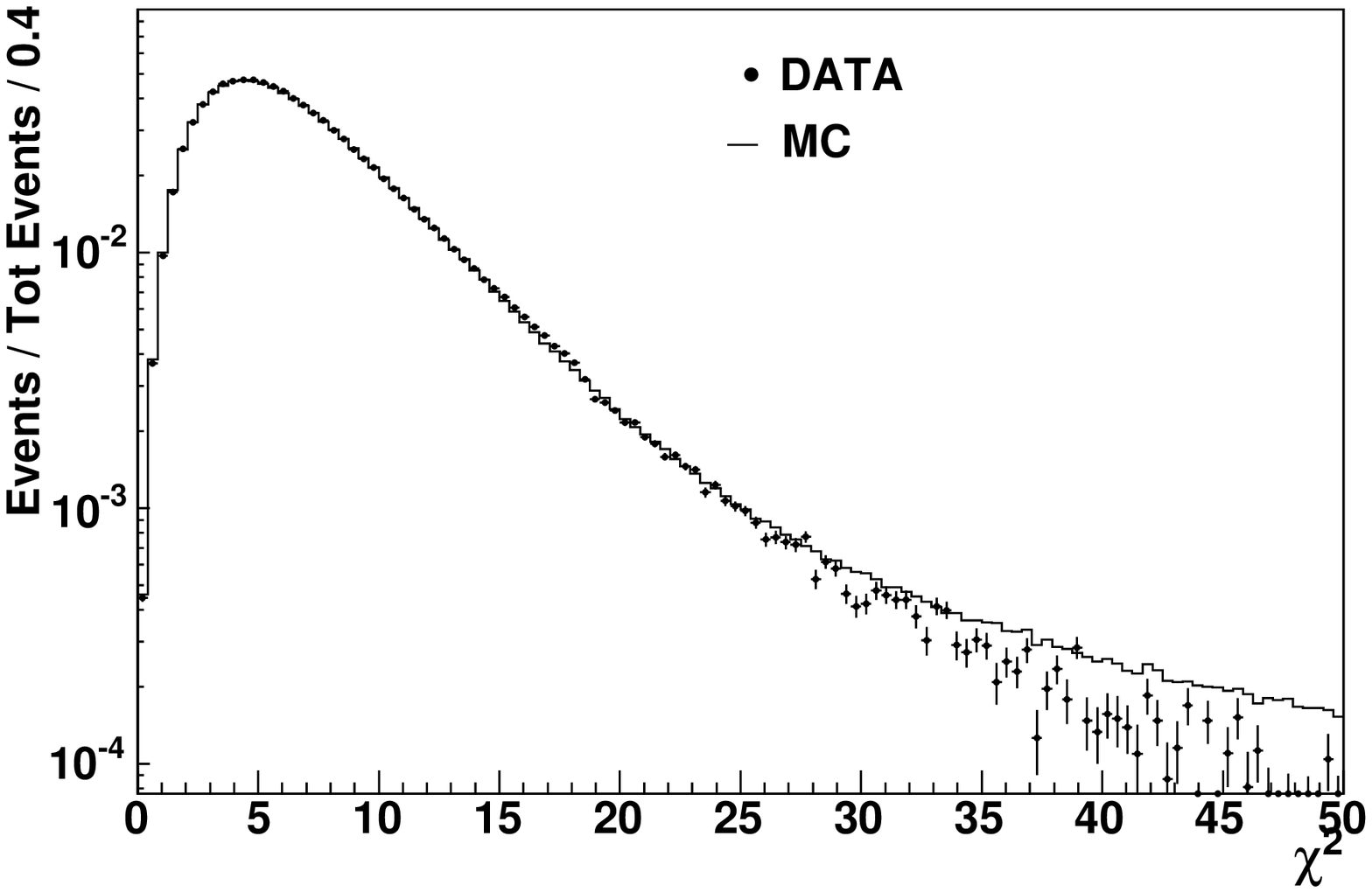, width=80mm}\\
\end{tabular}
\end{center}
\caption{{\footnotesize $\chi^2$ distribution for the kinematic fit. Left: linear scale; Right: log scale.}}
\label{chisq_eta_2}
\end{figure}
\item Finally we require:
\bi
\item  $320 \MeV < E_{\gamma rec} < 400 \MeV$ for the recoil photon
 (to reduce residual background from \fikskl events).
\item $E_{\pip}+E_{\pim}< 550 \MeV$ (to reduce residual background from \fipippimpiz events).
\item the invariant mass of the two softest photons: $M_{\gamma\gamma}$
  $\in [ 110, 160 ]$ MeV (to reduce residual background from 
  \etapippimpiz decays with $\piz \rightarrow e^{+}e^{-}\gamma$ and from
  \fietag events with $\eta \rightarrow \pip \pim \gamma$).\\
\ei
\end{enumerate}

The selection efficiency is determined by using the KLOE MonteCarlo (MC)
simulation program \cite{EVCL} and
checked on data by means of control samples. In particular:
\begin{itemize}
\item The trigger efficiency evaluated by MonteCarlo is $99.9$\%, and
 we have excellent  Data-MC agreement for the trigger sectors
 multiplicities.
\item The effect of the event classification procedure (EVCL) and
  machine background filter is evaluated using a downscaled 
  sample of non filtered data to which we apply much less stringent
  cuts  in order to
  have a ``minimum bias`` selection. On signal events the efficiency
  of the minimum bias selection is 99.88\%.\\
  We have found that the EVCL procedure
  introduces a signal loss of $\sim 1.5$\%,
  expected also from Monte Carlo estimates. The corresponding bias on the
  Dalitz plot parameters measurement has been evaluated and included in the
  systematic error.
  No bias is introduced by the FILFO procedure.\\
\item The tracking and vertex efficiencies have been estimated from
  the Data-MC ratio observed for a control sample of
  \fipippimpiz~ events selected to have charged-pion
  momenta in the same range as those from the \etapippimpiz~ decay \cite{CAMILLA_NOTE}:
  \begin{equation}
  \frac{\left(\varepsilon^{2}_{TRK}\varepsilon_{VTX}\right)_{Data}}{\left(\varepsilon^{2}_{TRK}\varepsilon_{VTX}\right)_{MC}} = 0.974~ \pm~ 0.006.
  \end{equation}
  \noindent
  The Data-MC ratio of efficiencies is flat all over
  the momentum spectrum, thus introducing no bias in the Dalitz plot
  fit. We recall that all the variables used in the fit are evaluated
  in the $\eta$ rest frame, which is boosted with respect to the
  laboratory by about 363 MeV/c momentum. Therefore to each momentum bin
  in the lab frame corresponds a much wider momentum interval 
  in the $\eta$ cms: any Data-MC
  discrepancy in the lab frame is further diluted by this effect.\\
\item A correction to the detection efficiency for low energy photons has been
  obtained by comparing the photon energy spectrum of a data subsample  to
  the expected MC spectrum; the average correction factor is
  0.964. 
\end{itemize}
\noindent
\\
The overall selection efficiency, taking into account all the Data-MC
corrections is found to be $\varepsilon = ( 33.4~ \pm~ 0.2 )$\%.
The expected background contamination, obtained from MC simulation is as low as 0.3\%. \\

After the background subtraction the observed number of events 
is: 
\begin{equation}
N_{obs} = 1.337 \pm 0.001~~~\textrm{Mevts}.\nonumber 
\end{equation}

The  distribution of the data in the Dalitz plot is shown in  fig.\ref{dalitz_plot}.

\bfig[h!]
 \begin{center}
  \epsfig{figure=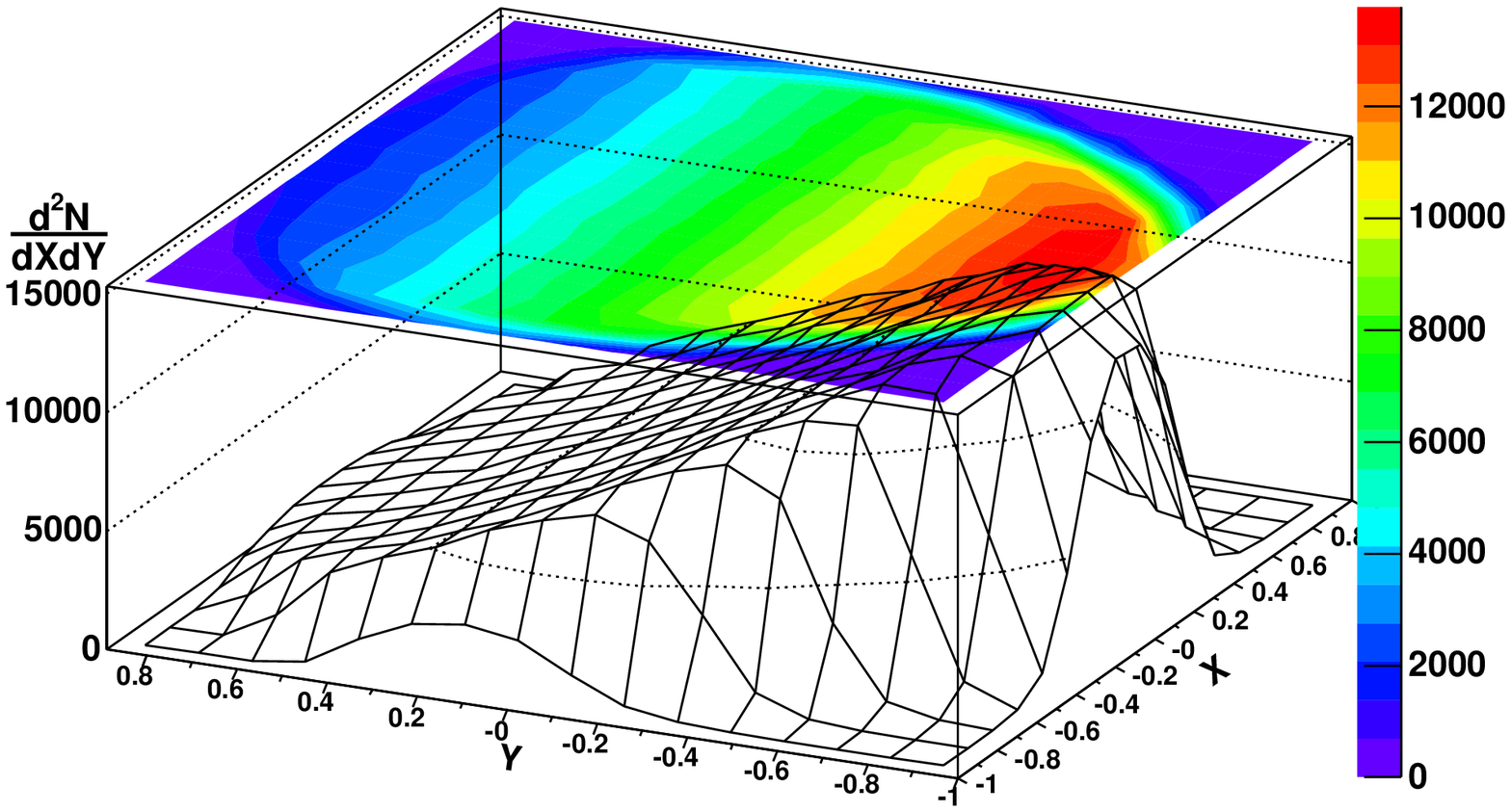,width=.8\linewidth}
  \caption{\footnotesize Dalitz-plot distribution for the whole
    data sample. The plot contains $1.34$ millions of events in $256$
    bins.} 
  \label{dalitz_plot} 
 \end{center}
\efig 

The signal selection efficiency $\varepsilon\left(X,Y\right)$ as function of Dalitz plot point
is obtained by MC, for each $\left(X,Y\right)$
bin, as the ratio: 
\begin{equation}
\varepsilon\left(X,Y\right) = \frac{N_{rec}\left(X,Y\right)}{N_{gen}\left(X,Y\right)}
\end{equation}
where $N_{rec}\left(X,Y\right)$ and $N_{gen}\left(X,Y\right)$ are respectively the
reconstructed and generated Dalitz distributions. 
This definition of efficiency takes into account the smearing effects due to the finite resolution in $X, Y$.
This approach is equivalent to the use of
the complete four-dimensional smearing matrix as long as  the
MC correctly reproduces the Dalitz plot shape.
To this aim we have used a  data subsample to obtain a first estimate
of the Dalitz plot parameters which we have used in final MC.

The efficiency $\varepsilon\left(X,Y\right)$ has a smooth behaviour all over the
Dalitz plot.
The projections of  $\varepsilon\left(X,Y\right)$ on the X and Y axis
are plotted in  fig.\ref{effXandY}.
\begin{figure}[h]
\begin{center}
\begin{tabular}{cc}
  \epsfig{figure=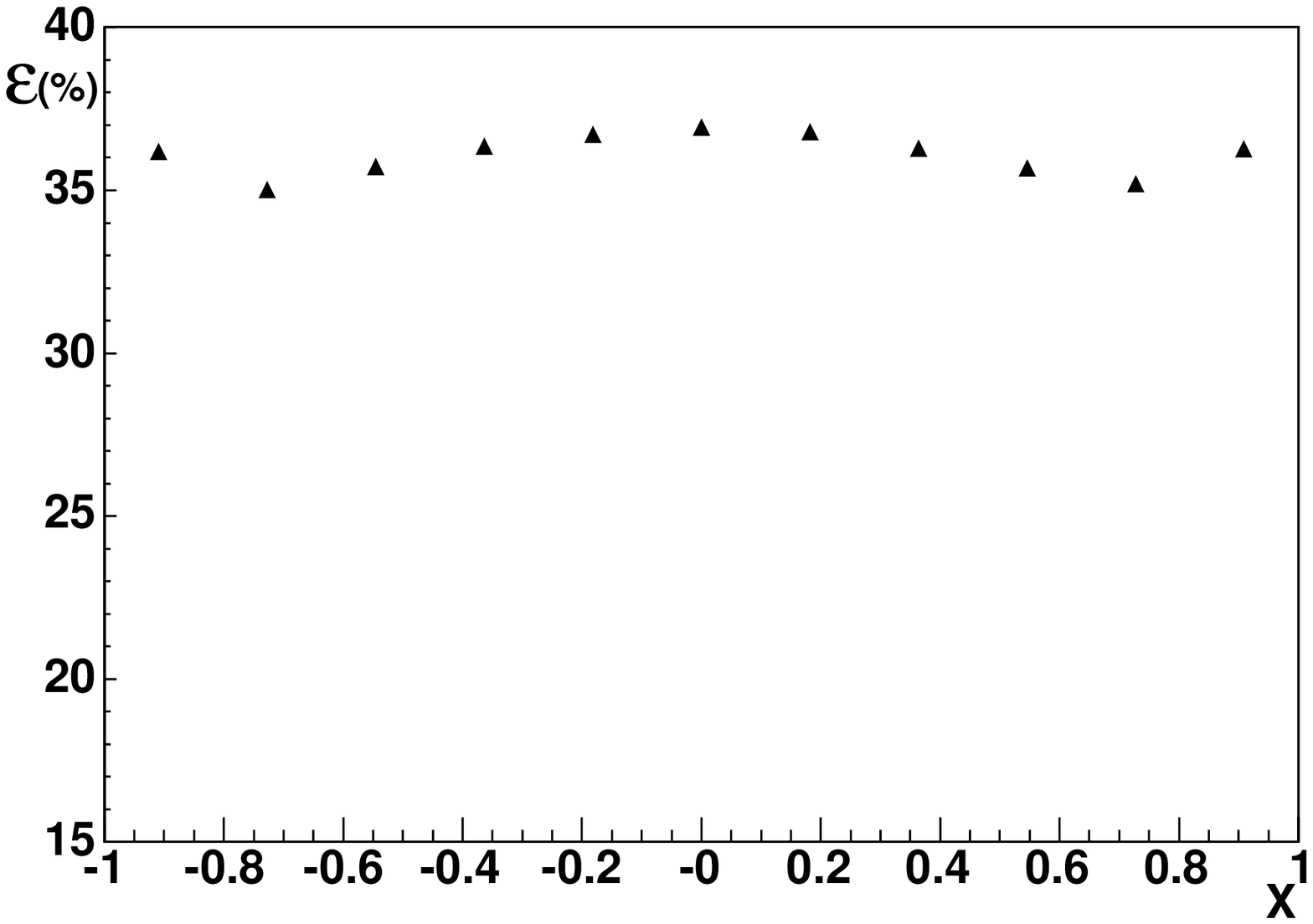,width=.5\linewidth}&
  \epsfig{figure=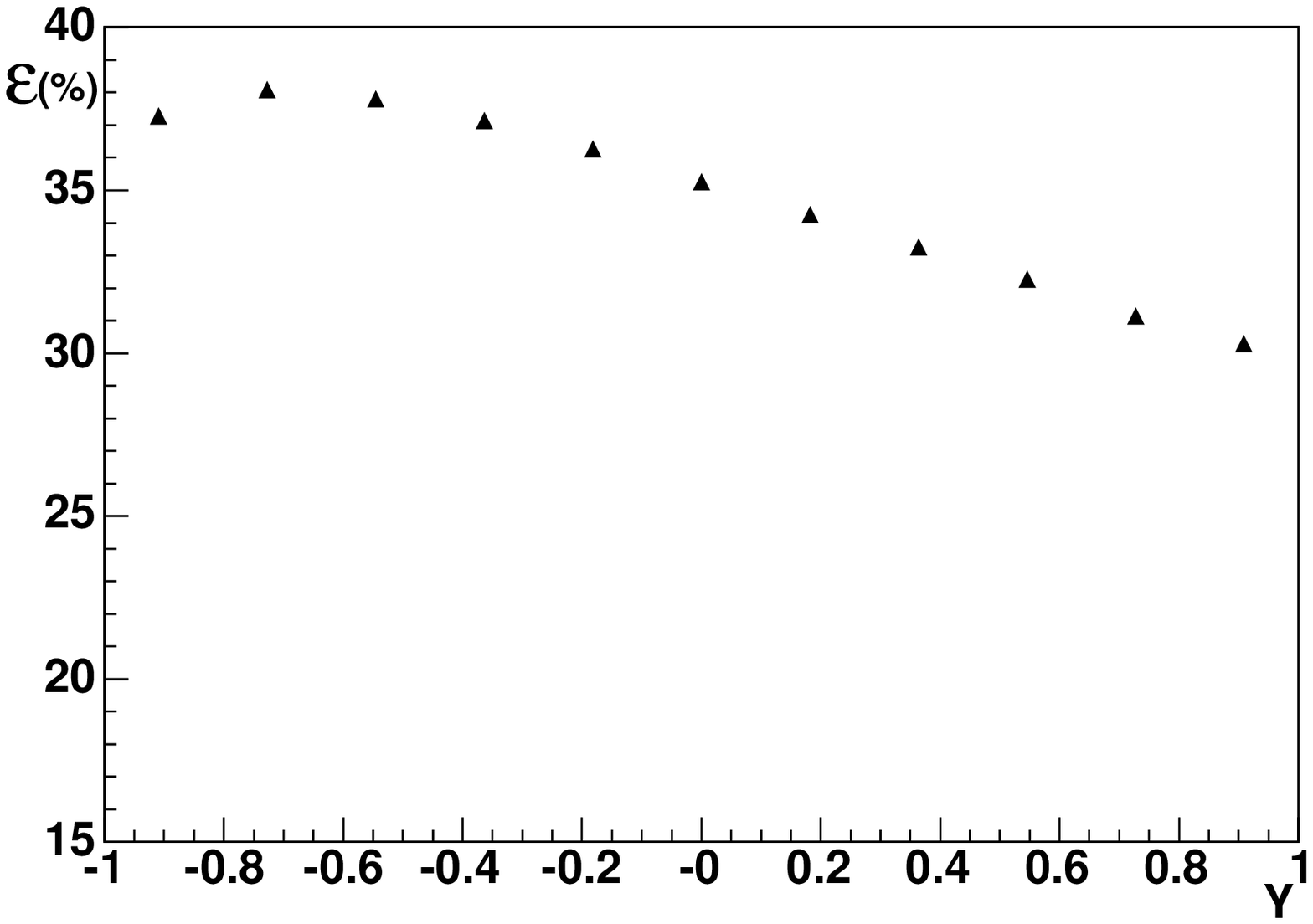,width=.5\linewidth}\\
\end{tabular}
\end{center}
\caption{{\footnotesize Left: Efficiency versus $X$. Right: Efficiency
    versus $Y$.}}
\label{effXandY}
\end{figure}
While the efficiency appears to be rather flat on $X$ (and
symmetric in X as expected), it decreases in an approximately
linear way with $Y$. 
In fact a large value for $Y$ means a low-momentum $\pi^{\pm}$
in the decay to which corresponds  a lower tracking/vertexing efficiency. 

The expected resolutions on the Dalitz variables $\left(X,Y\right)$ are  shown in
fig.\ref{resXandY}.
The core resolution on $X$  is about 0.02, due to our excellent momentum
resolution for charged tracks. The $Y$ variable, which is proportional to
the $\piz$ kinetic energy, is evaluated \cite{NOTA215} as the average between the
"direct" determination obtained from the energy and direction of the
two clusters associated to the $\piz\to\gamma\gamma$ decay and the
"indirect" determination :
$
T_{0} =  M_{\eta} - \left(E_{\pip}+E_{\pim}\right) - M_{\piz}
$. 
This  leads to a core resolution on $Y$ of about 0.02.

\begin{figure}[h]
\begin{center}
\begin{tabular}{cc}
\epsfig{figure=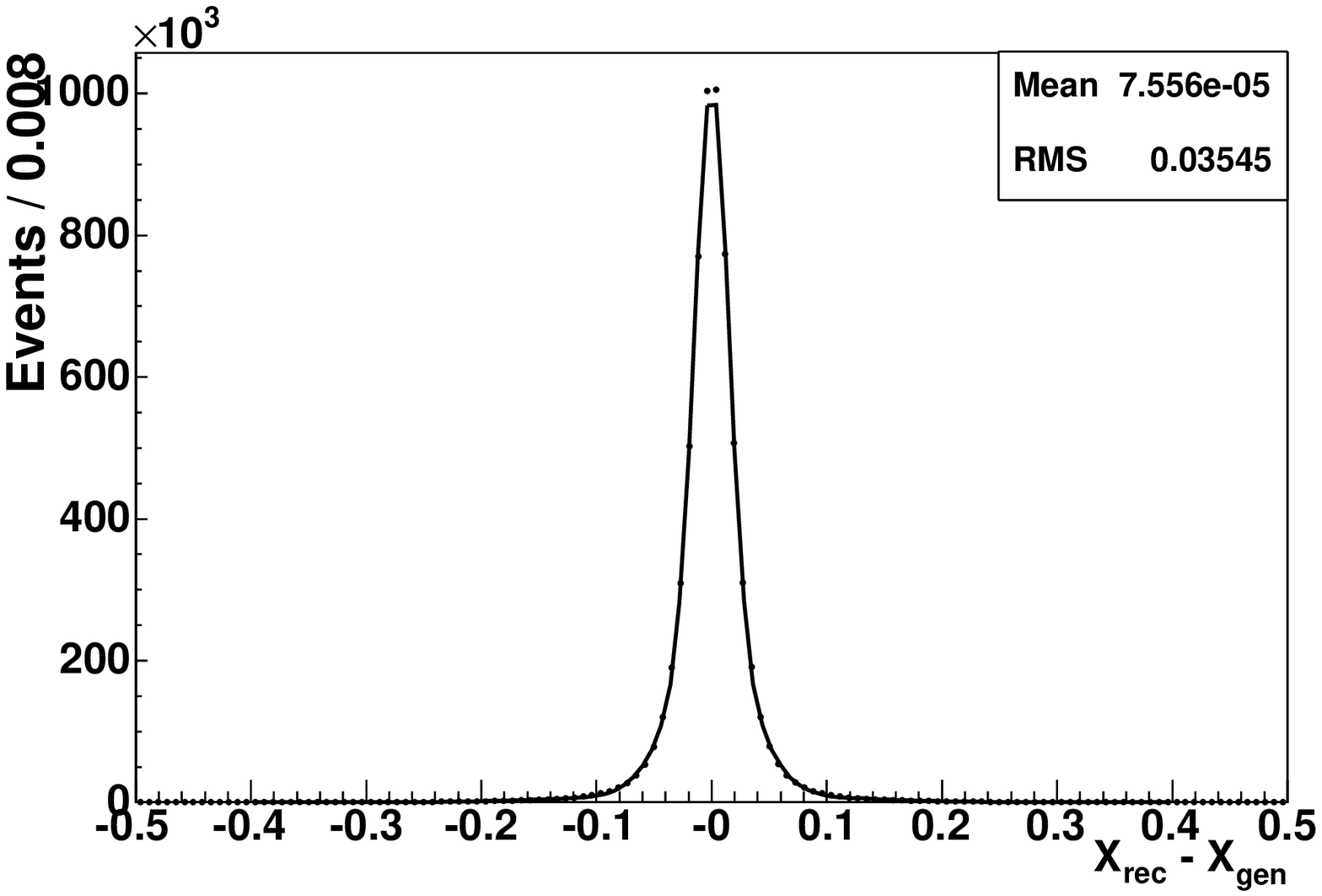,width=.5\linewidth }&
\epsfig{figure=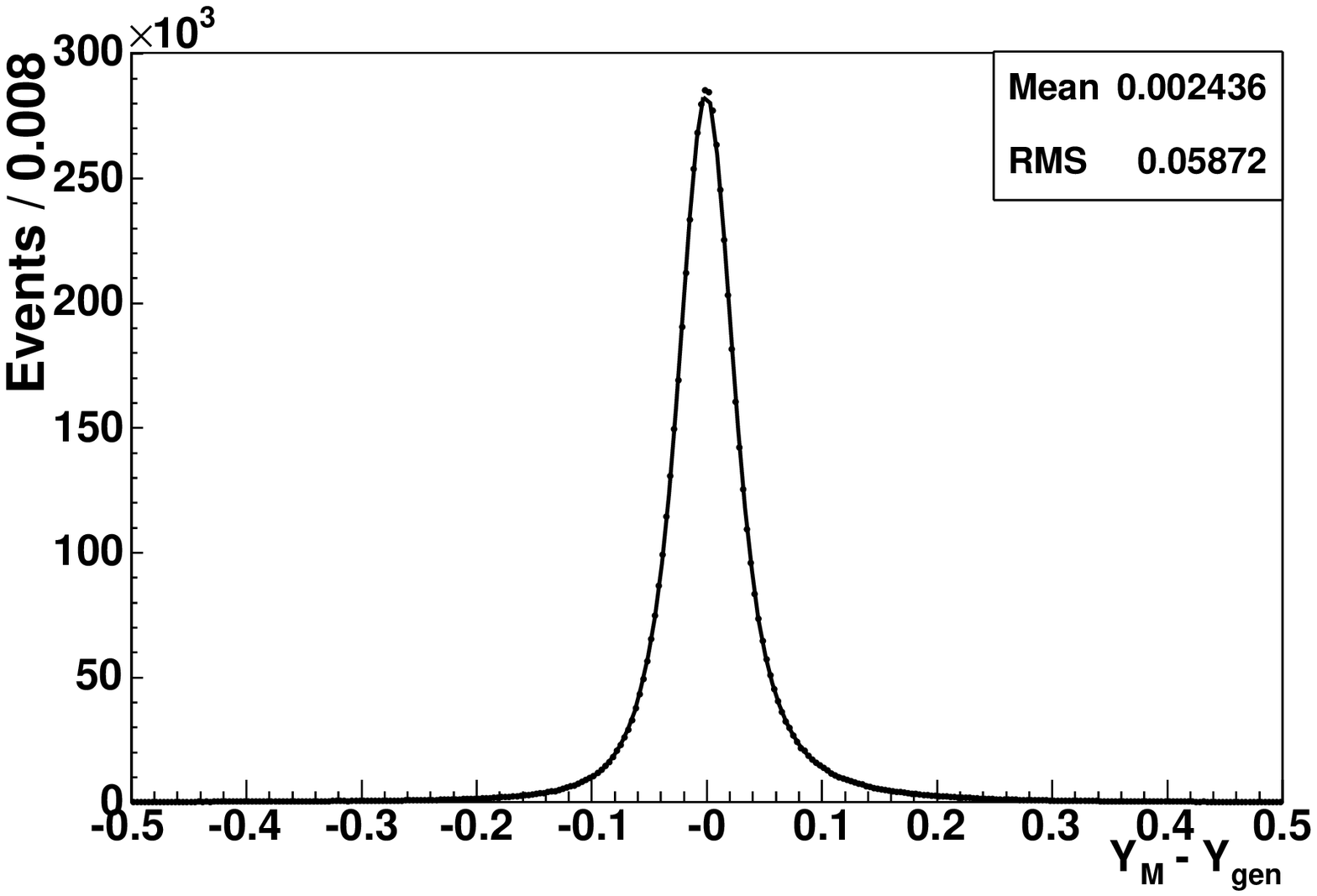,width=.5\linewidth }\\
\end{tabular}
\end{center}
\caption{{\footnotesize Resolutions on $X$
    (left) and $Y$ (right) according to MC. The curves are fitted to a sum of four
    gaussians.}}
\label{resXandY}
\end{figure}

\section{Fit of Dalitz plot}
The fit to the Dalitz plot is done using a least squares approach.
Let $\vert A(X,Y)\vert^{2}$ be the theoretical squared amplitude: 
\begin{equation}
\vert A(X,Y)\vert^{2} = N (1 + aY + bY^{2} + cX + dX^{2} + e XY + ...).
\label{eq:standardpar}
\end{equation}
\noindent
with N being a normalization constant.

Then the  $\chi^{2}$ is defined as: 
\begin{equation}
\chi^{2} = \sum_{i}\sum_{j}\left(
\frac{N_{ij}-\varepsilon_{ij} 
\int_{x_{i}}^{x_{i}+\Delta x}\int_{y_{j}}^{y_{j}+\Delta y} \vert
A(X,Y)\vert^{2} dPh(X,Y)}{\sigma_{ij}} \right)^{2} 
\label{eq:chi2}
\end{equation}
\noindent
where dPh(X,Y) is the phase space element and for each bin $(i,j)$:
\begin{itemize}
\item  $N_{ij}$ is the number of observed events in the bin,
\item $\varepsilon_{ij}$ is the signal selection efficiency,
\item $(x_{i},x_{i} +\Delta x)$ and $(y_{j},y_{j} +\Delta y)$ are
  the bin boundaries,
\item 
$\sigma_{ij} = \frac{N_{ij}}{\varepsilon_{ij}}\sqrt{\frac{1}{N_{ij}} + \left(\frac{
\left(\delta\varepsilon\right)_{ij}}{\varepsilon_{ij}}\right)^{2}}, 
$
\end{itemize}
\noindent
All the bins are included in the fit except those intersected by
the Dalitz plot contour. 
The fit procedure has been tested on Monte Carlo and has been verified to correctly  
reproduce in output the input values of the parameters.

When applying the fit on the full data set we observe that the
expansion up to the second order  does not fit adequately the data:
we find values of $\chi^{2}$ probability very low, for any choice of
the binning. Therefore we  added to the amplitude expansion  the cubic terms:
\begin{equation}
\vert A(X,Y)\vert^{2} \simeq 1 + aY + bY^{2} + cX + dX^{2} + eXY + fY^{3} + gX^{3} + hX^{2}Y + lXY^{2}.
\label{eq:nostandardpar}
\end{equation}
\noindent

The values of $\chi^{2}$ probability improve significantly with the
inclusion of only one additional term, namely the $f Y^3$ term; all the
other cubic terms are consistent with zero and do
not improve the fit quality; therefore in the following we keep only the
$f$ parameter.\\ 
Changing the bin size from  $\Delta X = \Delta Y =0.20$
to $\Delta X = \Delta Y =0.11$ , we find very similar results, showing
that the fit is almost binning independent.\\
In table \ref{tab:tot_stat} are reported
the fit results for $ \Delta X = \Delta Y =0.125$ (154 bins are used), where we found the
best $\chi^{2}$ probability, and for different parameterisations of
$\vert A\vert^{2}$.In the first row all the parameters $a,b,c,d,e,f$ are
left free; in the following rows the C-violating $c,e$ parameters are
always set =0; in the last 3 rows we have moreover set  $ d=0;~ f=0$ and $d=f=0.$\\

\begin{table}[!h]
  \vspace{+20 pt}  
  \begin{center}
  \begin{tabular}{|c|c|c|c|c|c|c|c|}
    \hline
    $dof$ & $P_{\chi^{2}}$ & 10 $^{3}$ $a$ & 10 $^{3}$ $b$ & 10 $^{3}$  $c$
    & 10 $^{3}$ $d$ & 10 $^{3}$ $e$ & 10 $^{3}$ $f$ \\
    \hline
    147 & 73\% & -1090$\pm$5 & 124$\pm$6 &
    2$\pm$3  & 57$\pm$6 & -6$\pm$7 & 140$\pm$10\\
    \hline
    149 & 74\% & -1090$\pm$5 & 124$\pm$6 &
    & 57$\pm$6 & & 140$\pm$10   \\
    \hline
    150 & $<10^{-6}$ & -1069$\pm$5 & 104$\pm$5 &
    &  & & 130$\pm$10 \\
    \hline
    150 & $<10^{-8}$ & -1041$\pm$3 & 145$\pm$6 &
    & 50$\pm$6  & &  \\
    \hline
    151 & $<10^{-6}$ & -1026$\pm$3 & 125$\pm$6 &
    &  & &  \\
    \hline
  \end{tabular}\\[2pt]
  \end{center}
  \caption{ Fit results for different parameterisations of $\vert
    A\vert^{2}$. Values in $2^{nd}$ row are used for final results.}
  \label{tab:tot_stat}
  \vspace{+20 pt}
\end{table}
As expected from C-invariance in \etapippimpiz decay the
parameters $c$ and $e$ are consistent with zero, and removing them
from the fit does not affect the result for remaining parameters.\\
We observe a quadratic slope in $X$ and a cubic slope in $Y$
clearly different from zero. 

Therefore our final results for the Dalitz plot parameters are those shown in second
row of the table.
Fig.\ref{agree_X_Y} shows a comparison between fit and data for the projections in $X$ or $Y$.\\
\begin{figure}[h]
\begin{center}
\begin{tabular}{cc}
  \epsfig{figure=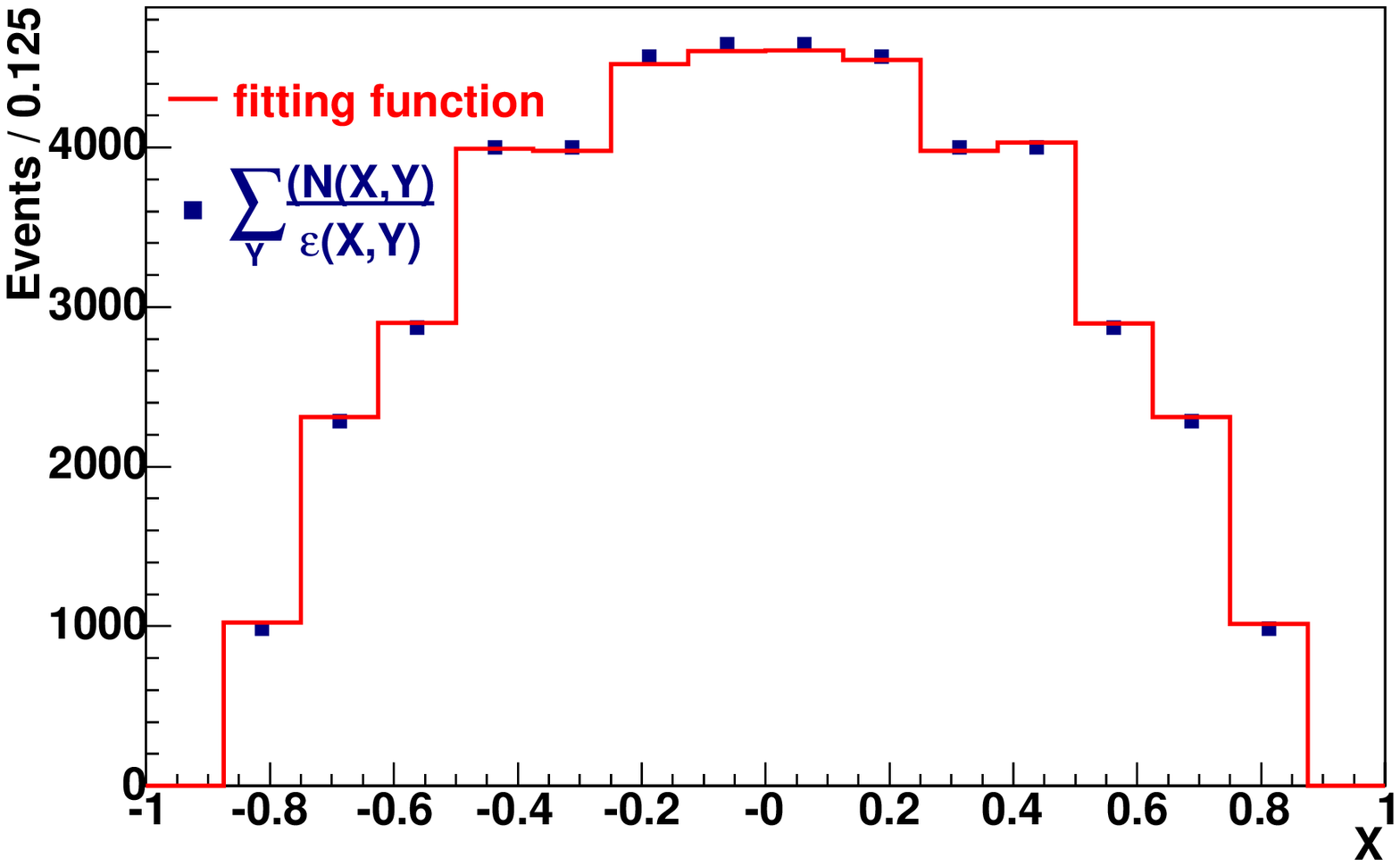,width=80mm}&
  \epsfig{figure=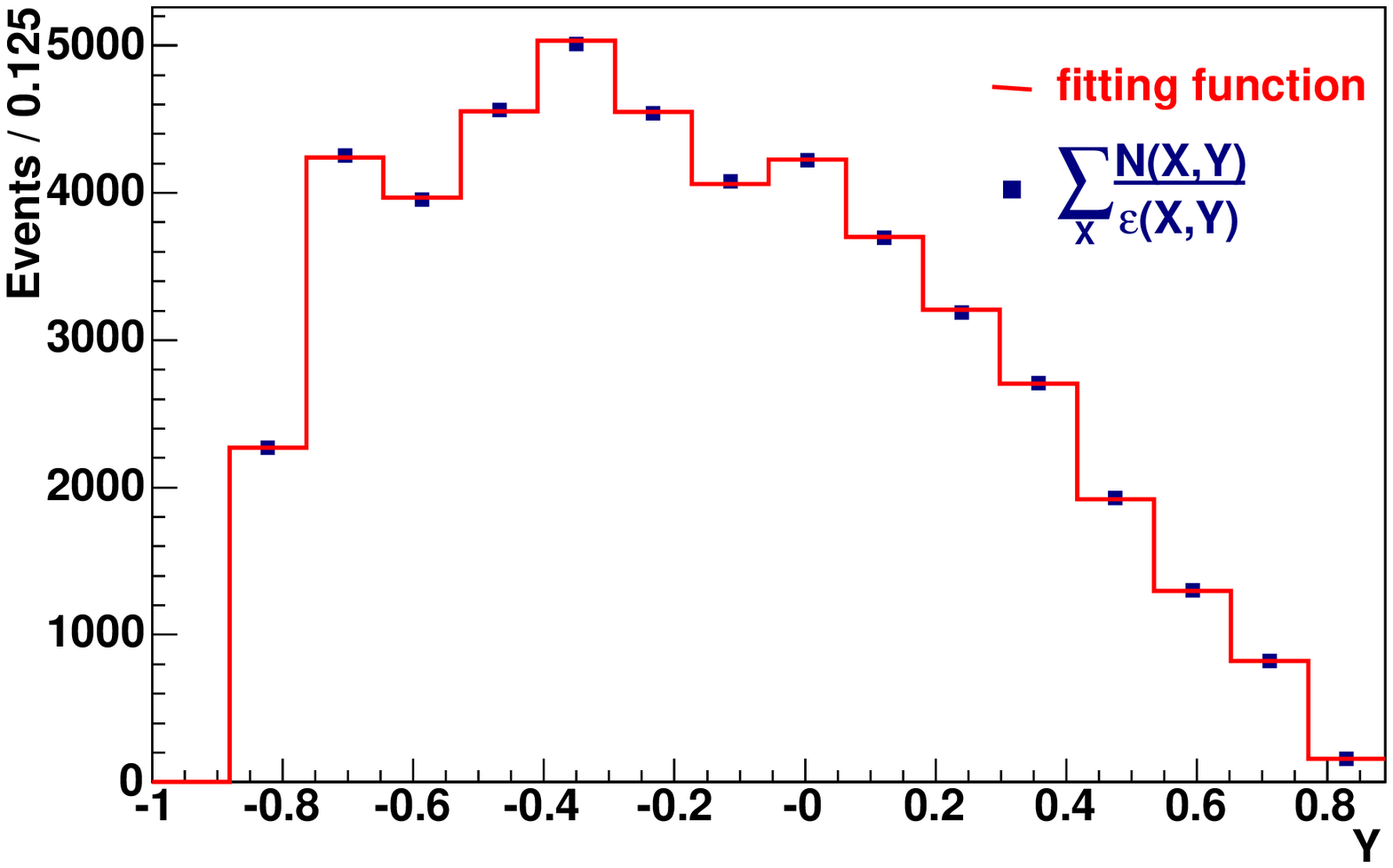,width=80mm}\\
\end{tabular}
\end{center}
\caption{{\footnotesize Comparison between data(points) and fit(histogram)
  for X,Y projections of the Dalitz plot distribution.}}
\label{agree_X_Y}
\end{figure}
\clearpage
\section{Systematic uncertainties \label{Systematics}}
We have estimated the systematics errors due to the following sources:

\begin{description}
\item[efficiency evaluation] 
All relevant reconstruction efficiencies have been checked directly on
data, using suitable control samples:
the only observable  systematic effect is introduced by the EVCL
procedure, which we have checked with the minimum bias sample.
We find a remarkable agreement between Data and MC for various kinematical
distributions (see fig.\ref{fig_pip_pim_fot_data_mc}); this implies that the
experimental resolution is well modelled in MC; therefore we neglect any
related systematics on the  signal selection efficiency
$\varepsilon\left(X,Y\right)$.
%
%
\item[resolution and binning] 
We have checked energy resolution for the photons by comparing the distributions  of
the photon energies after the kinematic fit on data and MC
finding good agreement for both the core and the tails of the
distributions. Drift Chamber momentum resolution and absolute scale is
controlled run by run by checking the $K_S$ mass reconstructed in
$\K_S\to\pi^+\pi^-$ events. The effect of binning was estimated by
changing the bin size by a factor of two in the
range $\Delta X = \Delta Y \in [ 0.11, 0.20 ]$.
\item[background contamination]
The main source of backgrounds to our analysis are:
\begin{enumerate}
\item $\fietag$ with $\etapippimpiz$ and $\piz\rightarrow e^{+}e^{-}\gamma$;
\item $\phi\rightarrow\omega\piz$ with $\omega\rightarrow\pip\pim\piz$;
\end{enumerate}
Although overall background contamination is small, its presence gives
an observable systematic effect at our level of precision.
We have changed the cut on $M_{\gamma\gamma}$ in a wide range,
corresponding to B/S varying from 0.7\% to 0.2\% , obtaining slightly
different values for the parameters.

\item[stability with respect to data taking conditions]
We have divided our data sample in 9 periods of about 50 \pbinv each. We have 
verified that the results for each parameter are compatible to be constant
all over the data taking and that the average values over the periods are
consistent with the ones from the whole data sample fit.
\item[radiative corrections]
Radiated effects have been considered by generating $10^7$
$\etapippimpiz\gamma$ decays, according to ref. \cite{Gatti}. We have verified that the bin by bin
ratio of the Dalitz plot obtained for
$\etapippimpiz\gamma$ decays with the one for $\etapippimpiz$ decays
can be fitted with a constant with $\chi^{2}/dof = 154/153$
corresponding to a $\chi^{2}$ probability of 46\%.\\
\end{description}
\noindent 
For each effect mentioned above the systematic error has been
conservatively estimated as the maximum parameter variation respect to 
 the reference one; the total systematic error is obtained by summing in
 quadrature the various contributions as shown  in table
\ref{tab:system_tot}.
%
 \begin{table}[!htb]
\vspace{+20 pt}  
   \begin{center}
     \begin{tabular}{|c|c|c|c|c|}
\hline
       Source &   $\Delta a$ & $\Delta b$ & $\Delta d$ & $\Delta f$ \\
       
\hline
       EVCL & -0.017 & 0.005 & -0.012 & 0.01  \\
\hline
       binning &  -0.008 +0.006 & -0.006 +0.006 & -0.007 +0.001 & -0.02 +0.02\\
\hline      
       background &  -0.001 +0.006 & -0.008 +0.006 & -0.007 +0.007 & -0.01 \\  
\hline
       Total  &  -0.019 +0.008 & $\pm$ 0.010 & -0.016 +0.007 & $\pm$ 0.02 \\  
\hline   
  \end{tabular}\\[2pt]
   \end{center}
   \caption{{\footnotesize Summary of
       the systematic errors on the Dalitz plot parameters.}}
   \label{tab:system_tot}
   \vspace{+20 pt}  
 \end{table}
\begin{figure}[htb]
\begin{center}
\begin{tabular}{ccc}
\epsfig{figure=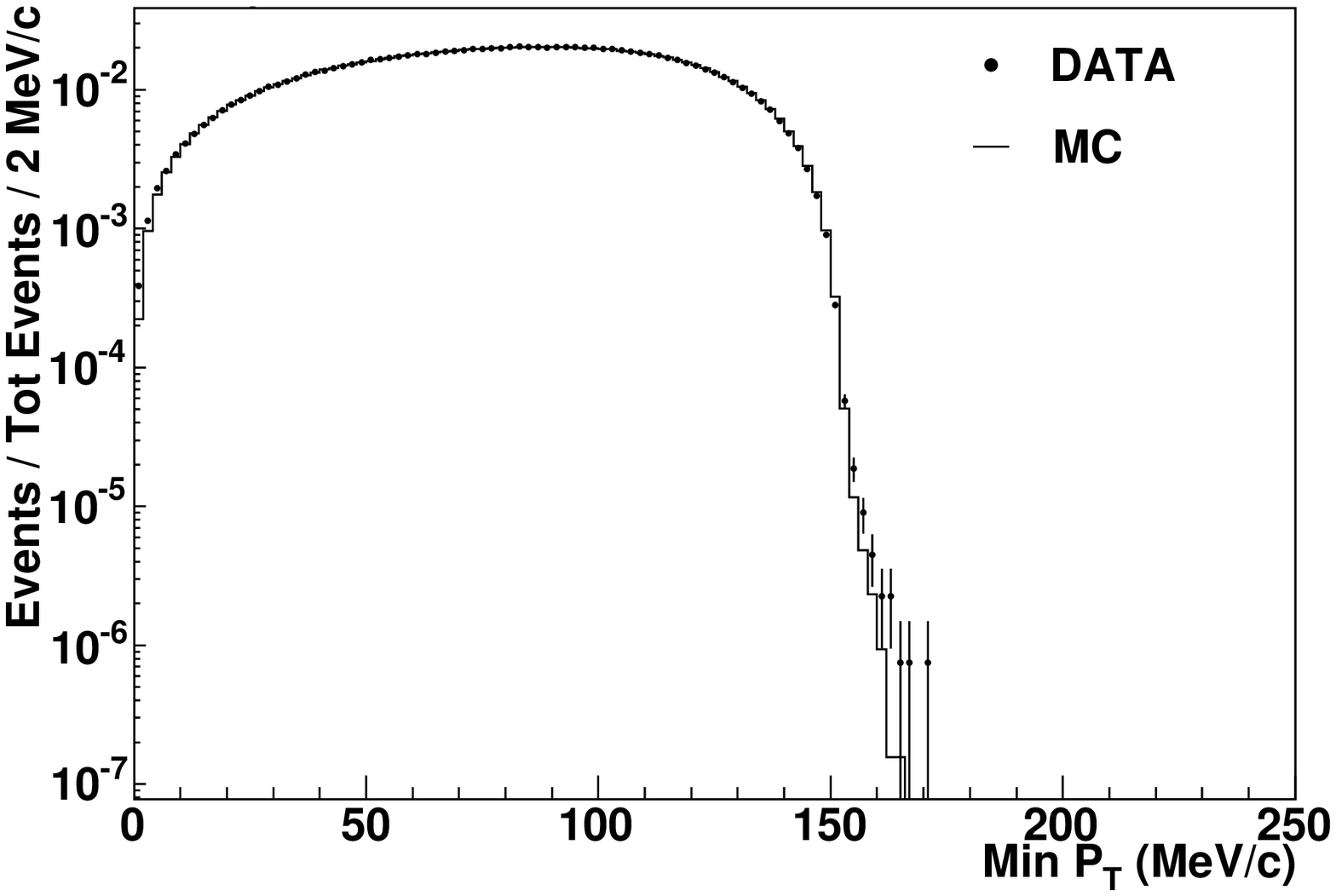,width=.5\linewidth}&
\epsfig{figure=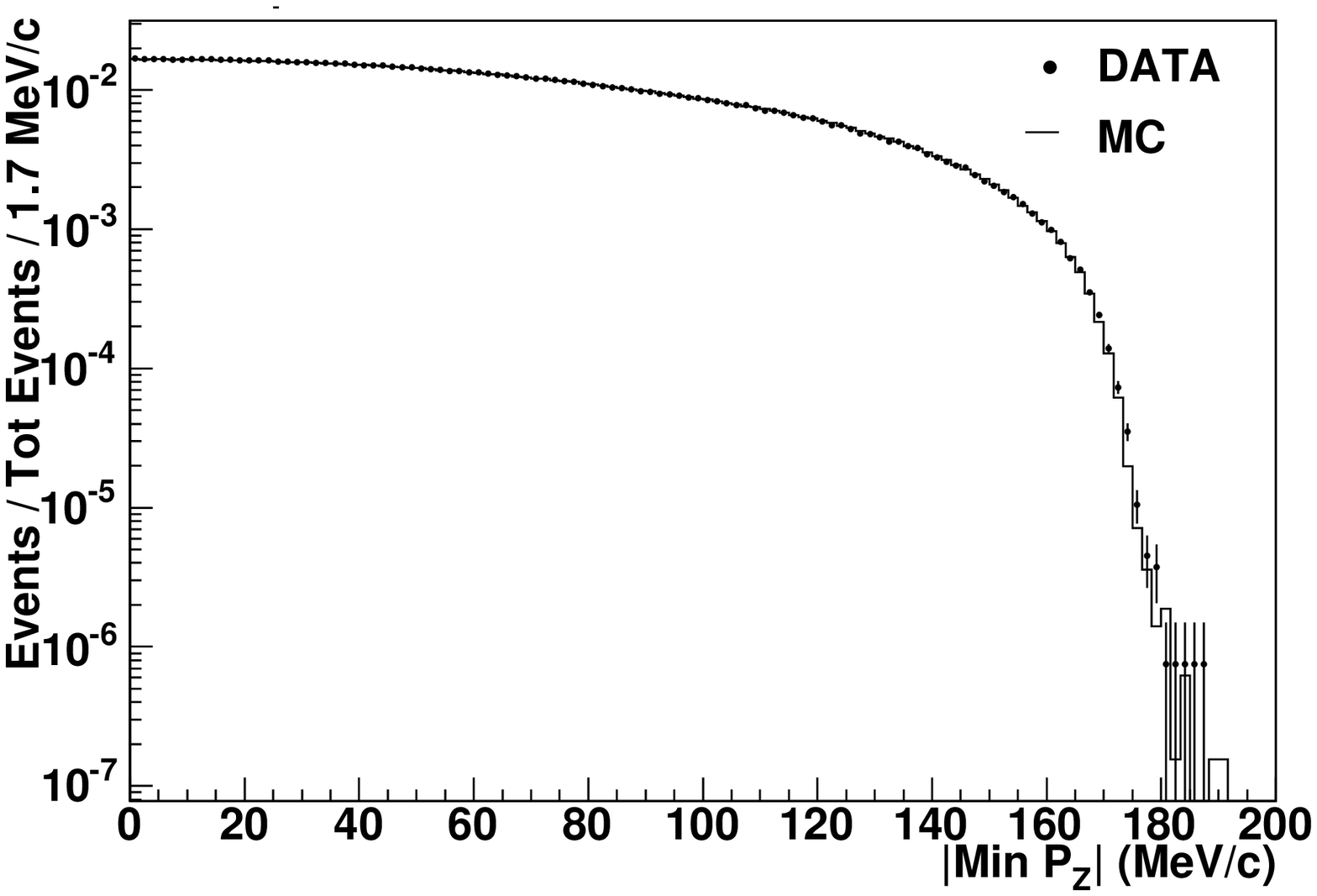,width=.5\linewidth}\\
\epsfig{figure=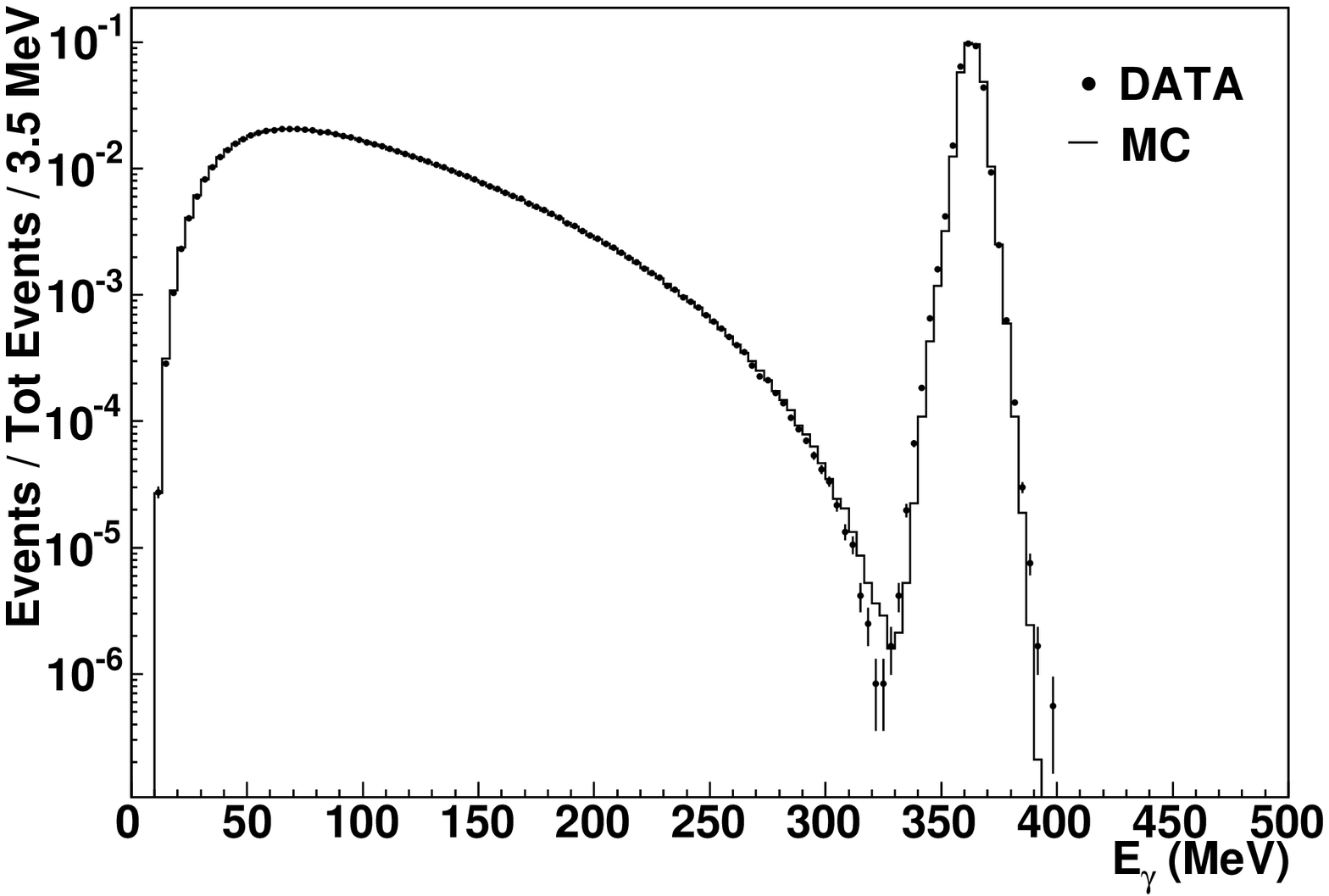,width=.5\linewidth}&
\epsfig{figure=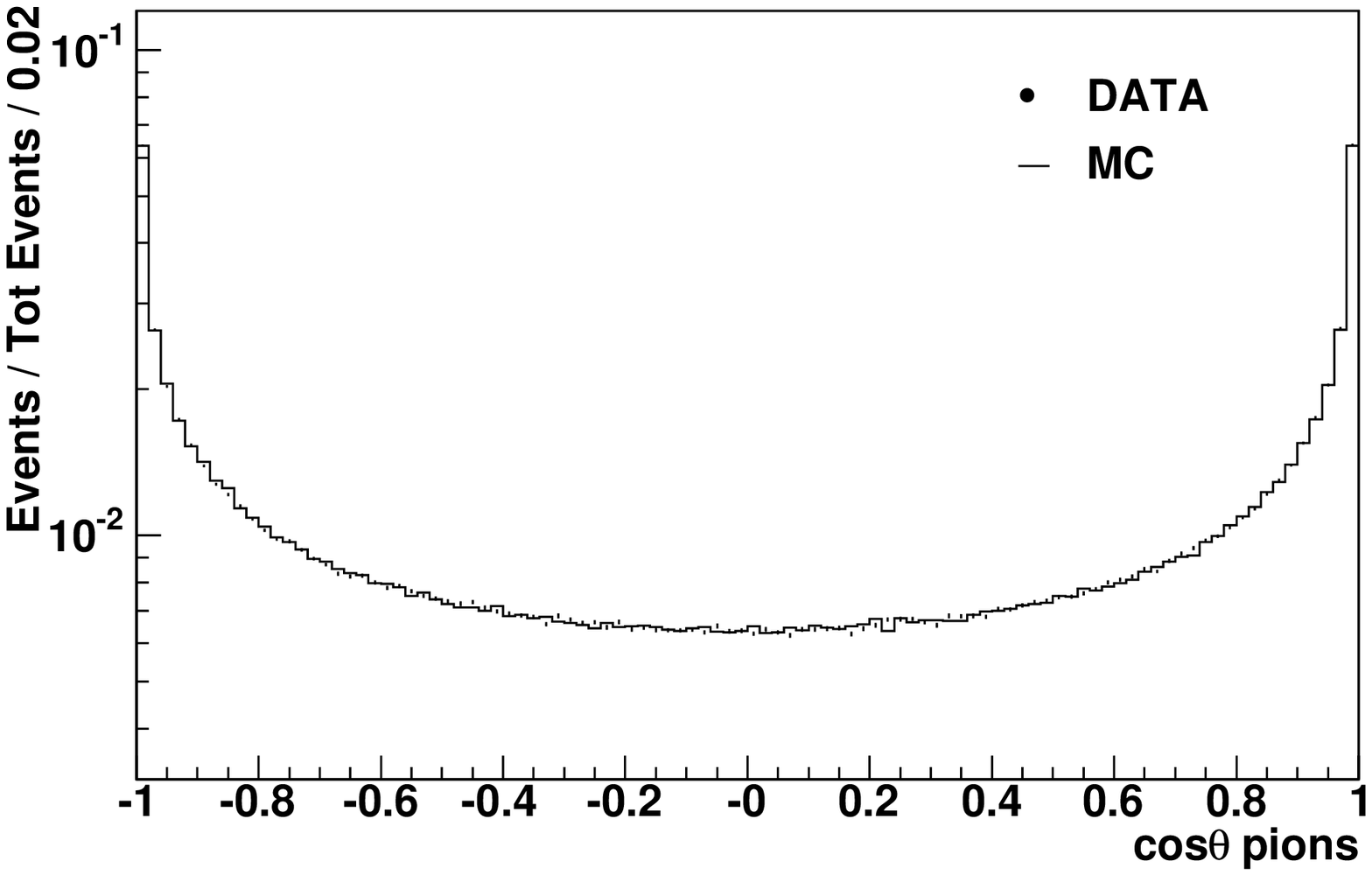,width=.5\linewidth}\\

\end{tabular}
\end{center}
\caption{{\footnotesize Data vs MonteCarlo comparisons in log scale. From top left in clockwise direction: minimum $P_T$ and $P_Z$, $\cos{\theta}$ between pion tracks and $E_{\gamma}$ for photons.}}
\label{fig_pip_pim_fot_data_mc}
\end{figure}
\section{Asymmetries}

While the polynomial fit of the Dalitz plot density gives valuable
information on the matrix element, some  integrated asymmetries are
very sensitive in assessing the possible contributions to C violation
in amplitudes with fixed $\Delta I$. 
In particular left-right asymmetry (which is of course strongly
related to the $c$ parameter in our fit) tests C violation with no
specific $\Delta I$ constraint; 
quadrants asymmetry tests C violation in $\Delta I = 2$ and sextants
asymmetry (for a  definition see ref. \cite{Layter}) 
 tests C violation in $\Delta I = 1$. 

In the following we present results on asymmetries which use 4 times
 the statistics entering the PDG fits.
For this measurement we use the following approach: 
we obtain from MC the efficiency, for each region of the Dalitz plot,
as the ratio between reconstructed and generated events in the region.
This definition takes into account the resolution effects as well. We cross
check the result by evaluating the asymmetries on Monte Carlo, these
turn out to be all compatible with zero. We then evaluate the
asymmetry on data by folding in the expected MC efficiency. 
after  subtracting  the MC expected background.
On MC, for a sample of $5.69\times 10^{6}$ events we get:
\bigskip
\begin{tabular}{ccc}
$\varepsilon_L = (34.91\pm 0.02)\%$ & $\varepsilon_R = (35.05\pm0.02)\%$ &$ A_{LR} = (-0.006 \pm 0.06 )\times 10^{-2}$\\
$\varepsilon_{13} = (35.01\pm 0.02)\%$ & $\varepsilon_{24} = (34.95\pm0.02)\%$&$ A_{Q}  = ( -0.008 \pm 0.06 )\times 10^{-2}$\\
$\varepsilon_{135} = (35.00\pm 0.02)\%$ &$\varepsilon_{246} = (34.96\pm0.02)\%$&$ A_{S}  = (-0.05 \pm 0.06 )\times 10^{-2}$\\
\end{tabular}
\\
The "raw" asymmetries on data are found to be:
$$
A_{LR} = (-9 \pm 9 )\times 10^{-4};\\
A_{Q}  = (2 \pm 9 )\times 10^{-4};\\
A_{S}  = (13 \pm 9 )\times 10^{-4}.\\
$$
Correcting for MC efficiencies we get:
$$
A_{LR} = (9 \pm 10_{stat.} )\times 10^{-4};\\
A_{Q}  = (-5 \pm 10_{stat.} )\times 10^{-4};\\
A_{S}  = (8 \pm 10_{stat.} )\times 10^{-4}.\\
$$

In assessing the systematics for the measured asymmetries we have
considered all the effects taken into account for the Dalitz plot fit,
namely: 
\begin{itemize}
\item Effect of background (by varying cuts);
\item Effect of EVCL (by using Minimum Bias sample);
\item Data-MC efficiency comparison (using the $\phi \to
  \pi^+\pi^-\pi^0$ control sample)
\end{itemize}
In particular the tracking efficiency has been evaluated separately
for the two charges, since in MC  is evident a small
but statistically significant difference in left and right
efficiencies, due to a slightly different tracking efficiency as
a function of $p_T$ for positive and negative pions.

Since we require both tracks to be reconstructed
the absolute value of the efficiency is not important for the
asymmetry, but rather its dependence upon the pion momentum. The very
good Data-MC agreement has been already demonstrated for both charges
on the signal. We here use an independent control sample of $\phi \to
\pi^+\pi^-\pi^0$ events and, as described before (see section
\ref{Efficiencies}), check the agreement between Data and MC for the
efficiencies as a function of momentum the two charges. Results are
shown in fig. \ref{Eff_charges}. The control sample agrees well
with MC within errors, and the Data-MC ratio is well fitted by a
constant. 
\begin{figure}[h]
\begin{center}
\epsfig{figure=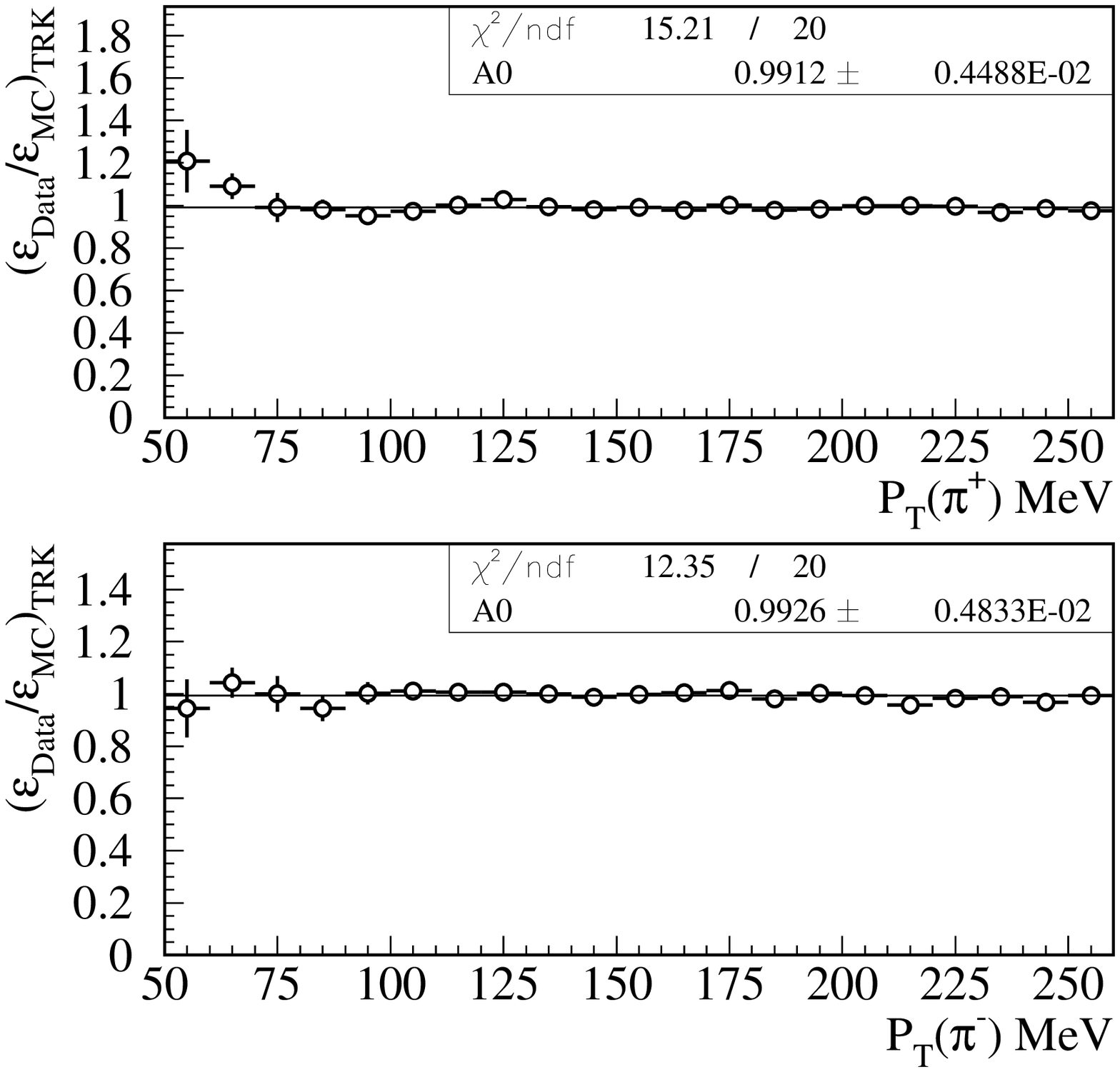,width=100mm}
\end{center}
\caption{{\footnotesize The Data-MC ratio of tracking efficiency for
    $\pi^+$ (up) and $\pi^-$ (down) as a function of the pion
    transverse momentum. The maximum deviation from a constant fit has been
    used for systematics evaluation (see text).}} 
\label{Eff_charges}
\end{figure}
In order to assess the possible systematics connected with the tracking
efficiencies we have assumed a ``worst case'' approach: we have estimated
the maximum positive or negative linear slopes compatible within one sigma with the
fit of the distributions shown in fig. \ref{Eff_charges}. Then we
have assumed the two charges to behave with  {\em opposite}
slopes. This gives us two possibilities ($\pi^+$ with positive slope
and $\pi^-$ with negative slope and vice-versa). We have then
reweighted the events according to these two possibilities and used
the maximum difference observed in the asymmetries as an estimate of
the possible systematic effects due to different Data-MC
efficiencies. 
The systematics connected with the
asymmetries are shown in table \ref{tab:asym_tot}.

 \begin{table}[!htb]
\vspace{+20 pt}  
   \begin{center}
     \begin{tabular}{|l|c|c|c|}
       \hline
       Syst. Effect &Left-Right &    Quadrant &  Sextant \\
       \hline
       Background & $(-0.2 / +0.1)\times 10^{-3}$ & $(-0.2/ +0.2)\times 10^{-3}$ &$( +0.3)\times 10^{-3}$\\
       EVCL & $(-0.5 )\times 10^{-3}$ & $(-0.3)\times 10^{-3}$ &$( +0.7)\times 10^{-3}$\\
       Efficiency&  $(-1.3 / +0.9)\times 10^{-3}$ & $(-0.3/ +0.2)\times 10^{-3}$ &$( -1.3)\times 10^{-3}$\\
       Total & $(-1.4 / +0.9)\times 10^{-3}$ & $(-0.5/ +0.3)\times 10^{-3}$ &$( -1.3/+0.8)\times 10^{-3}$\\
       \hline
     \end{tabular}\\[2pt]
   \end{center}
   \caption{\footnotesize{Systematic errors on asymmetries. }}
   \label{tab:asym_tot}
   \vspace{+20 pt}
 \end{table}

Therefore the final results for the asymmetries are:
$$
A_{LR} = (0.09 \pm 0.10 (\rm stat.)^{+0.09}_{-0.14} (\rm syst.) )\times 10^{-2}
$$
$$
A_{Q}  = ( -0.05 \pm 0.10 (\rm stat.)^{+0.03}_{-0.05} (\rm syst.) )\times 10^{-2}
$$
$$
A_{S}  = (0.08 \pm 0.10 (\rm stat.)^{+0.08}_{-0.13} (\rm syst.) )\times 10^{-2}
$$
\section{Conclusions}
The results including the statistical uncertainties coming from the
fit and the estimate of systematics are:
\begin{equation}
a = -1.090 \pm 0.005 (stat) ^{+ 0.008}_{- 0.019} (syst)
\end{equation}
\begin{equation}
b = 0.124 \pm 0.006 (stat) \pm 0.010 (syst)
\end{equation}
\begin{equation}
d = 0.057 \pm 0.006 (stat) ^{+ 0.007}_{- 0.016} (syst)
\end{equation}
\begin{equation}
f = 0.14 \pm 0.01 (stat) \pm 0.02 (syst)
\end{equation}
The systematic error has been obtained adding in quadrature
all the contributions in tab. \ref{tab:system_tot}. 
Tab.\ref{tab:matrix_corr_fin} gives the correlation coefficients
between the fitted parameters.\\ 
\begin{table}[h]
  \begin{center}
    \begin{tabular}{|c|c|c|c|c|}
      \hline
      & $a$ & $b$ & $d$ & $f$ \\
      \hline 
      $a$        & 1     & -0.226 & -0.405 & -0.795 \\ 
      $b$        &       & 1      &  0.358 &  0.261  \\ 
      $d$        &       &        &  1     &  0.113 \\  
      $f$        &       &        &        & 1      \\ 
      \hline
    \end{tabular}
  \end{center}
  \caption{\footnotesize Error matrix from the Dalitz plot fit.}  
  \label{tab:matrix_corr_fin}
\end{table}
\noindent
The following comments can be done:
\begin{itemize}
\item the fitted value for the quadratic slope in $Y$ is almost one half of
  the simple Current Algebra prediction ($b = a^{2}/4$), thus calling for
  significant higher order corrections;
\item the quadratic term in $X$ is unambiguously different from
  zero; 
\item similarly for the large cubic term in $Y$;

\item when integrating the polynomial over the phase space to get the decay
 width, the strong correlations between parameters must be carefully taken into
 account for a  correct error  estimate.

\item we don't observe any evidence for $C$ violation in
the $\eta\to\pip\pim\piz$ decay since the $c$ and $e$ parameters of the
Dalitz plot and the charge
asymmetries are all well consistent with zero.

\end{itemize}

\end{document}